\RequirePackage[2020-02-02]{latexrelease}
\documentclass[
aps,%
12pt,%
final,%
notitlepage,%
oneside,%
onecolumn,%
nobibnotes,%
nofootinbib,
superscriptaddress,%
noshowpacs,%
centertags]%
{revtex4}

\usepackage{graphicx}
\usepackage{dcolumn}
\usepackage{bm}
\usepackage[english]{babel}
\usepackage{amsmath}
\makeatletter
\renewcommand{\@biblabel}[1]{#1.} 
\makeatother

\begin{document}
\selectlanguage{english} 


\title{Application of the Holographic Equation of State for Numerical Modeling of the Evolution of Quark-Gluon Plasma}
\author{\firstname{Anton} \surname{Anufriev}}
\email[]{anton.anufriev@spbu.ru}
\affiliation{Saint Petersburg State University}
\author{\firstname{Vladimir} \surname{Kovalenko}}
\email[]{v.kovalenko@spbu.ru}
\affiliation{Saint Petersburg State University}

\begin{abstract}

In this paper, we propose a method for numerical modeling of the nuclear matter properties within the framework of relativistic heavy-ion collisions using a holographic equation of state. Model's free parameters are calibrated using lattice results for quark masses approximating physical values and  adjusted to match the Regge spectra of $\rho$ mesons. Numerical simulations are performed using the iEBE-MUSIC framework, which incorporates the MUSIC relativistic hydrodynamics solver. We modify the code by implementing a tabulated holographic equation of state, enabling simulations of quark-gluon plasma evolution with dynamically generated initial conditions via the 3D Monte Carlo Glauber Model. Finally, the spectra of produced hadrons are computed using a hybrid iSS+UrQMD approach at the freeze-out stage.
\end{abstract}

\begingroup
\makeatletter
\renewcommand{\@fnsymbol}[1]{
  \ifcase#1\relax    
  \or *
  \or **
  \or ***
  \or ****
  \else\@ctrerr\fi   
}
\renewcommand{\thefootnote}{\fnsymbol{footnote}} 
\maketitle
\endgroup

\section{Introduction}
The multiparticle dynamics of Quantum Chromodynamics (QCD) at high baryonic matter densities encompasses a broad range of fundamental physics problems, including deconfinement phenomena. Since the 1980s, lattice QCD calculations have indicated the existence of a phase transition when thermodynamic parameters reach critical values. This transition is linked to the formation of a new state of deconfined matter: quark-gluon plasma (QGP).~\cite{1}.

For decades, hydrodynamic models have been successfully used to study the evolution of quark-gluon plasma (QGP), proving highly effective in describing multiparticle production~\cite{2}. The dynamics of a relativistic fluid is fully determined by its velocity vector, energy-momentum density tensor, and baryon density, taking into account an equation of state (EoS) of the medium. Identifying the most accurate form of this equation remains a crucial challenge. For instance, lattice QCD at zero baryon chemical potential $\mu_B$ facilitates the calculation of the EoS as a function of temperature, revealing the smooth crossover nature of the QCD phase transition~\cite{3}. However,  at nonzero $\mu_B$ the ``sign problem'' arises due to the non-positive-definite nature of the fermionic determinant~\cite{4}.  This complication---often linked to the potential existence of a critical point in the QCD phase diagram---has motivated the exploration of more exotic approaches for the EoS.

Following the emergence of AdS/CFT correspondence in 1998~\cite{5}, numerous studies have concentrated on the prospective relevance of the approach based on the holographic duality of quantum chromodynamics (QCD) and gravitational theory in the anti-de Sitter space (AdS/QCD duality)~\cite{6}. 
In the research~\cite{7}, the authors introduce an additional scalar dilaton field within the classical Einstein-Maxwell action and apply the so-called potential reconstruction method. Free parameters appear in the deformed initial ansatz of gravitational theory, and a dilaton potential is calculated from the equations of motion.  That paper introduces the  ``light quarks'' model, which effectively reproduces the behavior of observables from lattice calculations in the chiral limit of light quark mass $m_q\to 0$ by adjusting the corresponding model parameters. This line of research was further pursued by the I. Ya Arefieva's group. The work~\cite{8} represents an effort to reconcile existing holographic models with the favorable outcomes of employing the anisotropic model of QGP holographic thermalization to analyze the dependence of the experimentally observed multiplicity density on the collision energy~\cite{9}.

It is evident that most of these holographic studies remain confined to theoretical investigations and rarely extend to analyzing available experimental data or formulating future predictions. This situation is a key motivating factor for applying holography in numerical simulations of QGP evolution within the framework of relativistic heavy-ion collisions.

\section{The holographic approach applied}
In this research, EoS of the QGP will be obtained by employing a special type ``bottom-up'' soft-wall (SW) approach. This connects the quasi-conformal QCD theory with classical gravity in the AdS space of dimension 5, as proposed in~\cite{8}. An action incorporating simultaneously two dilaton fields of the following form is introduced:
\begin{eqnarray}\label{eq1}
 S=\frac{1}{16\pi G_5}\int d^5x \sqrt{-\det(g_{\mu\nu})} \left[R-\frac{f_1(\phi)}{4}F_{(1)}^2-\frac{f_2(\phi)}{4}F_{(2)}^2-\frac{1}{2}\partial_{\mu}\phi \partial^{\mu}\phi-V(\phi)  \right],   
\end{eqnarray}
where $\phi$ is a scalar dilaton field, its potential of unknown type is $V(\phi)$, $F_{MN}^{(a)}=\partial_MA_N^{(a)}-\partial_NA_M^{(a)}$, $f_i(\phi)$ are functions of interaction with Maxwell fields, $i = 1,2$, 
$\det(g_{\mu\nu})$ is the determinant of the metric, and $G_5$ is the gravitational constant in AdS$^5$.

 A special form of the ansatz is introduced in addition to (1):
\begin{eqnarray}\label{eq2}
ds^2=\frac{R^2}{z^2}b(z)\left[ -g(z)dt^2+dx^2+\left(\frac{z}{R}\right)^{2-\frac{2}{\nu}}dy_1^2+\left(\frac{z}{R}\right)^{2-\frac{2}{\nu}}dy_2^2+\frac{dz^2}{g(z)} \right],
\end{eqnarray}
where $R$ is a dimensional factor that corresponds to the AdS radius for the standard Poincare metric ($R=1$ will be assumed for further calculations. This is a typical choice for theoretical works and does not reduce the generality of the above expressions), $g(z)$ is the blackening function that determines the thermodynamic behavior of the black brane. The parameter $\nu$ controls the spatial anisotropy of the metric components, 
where $\nu=1$ 
corresponds to the isotropic case.

The deformation factor $b(z)$ from Eq. (\ref{eq2}) in the Ref.~\cite{8} corresponds to the ``light quarks'' model and was chosen in such a way as to restore the results of lattice QCD calculations in the chiral limit. Thus:
\begin{eqnarray} \label{eq3}
b(z)=\exp({2A(z)}), \\
A(z)=-a\ln(bz^2+1). \label{eq4}
\end{eqnarray}

The expressions corresponding to the thermodynamics of a black hole were obtained by solving the equations of motion from Eqs. (\ref{eq1}) and (\ref{eq2}) with (\ref{eq3}). Taking into account the unit assignment of the constant $L$, determined by the characteristic radius AdS$^5$ $L=1$, thermodynamic parameters of QGP will take the following form:
 \begin{eqnarray}
T=\frac{1}{4\pi}\left| -\frac{(1+bz_h^2)^{3a}z_h^{1+\frac{2}{\nu}}}{I_1}\left[ 1-\frac{2\mu^2ce^{cz_h^2}}{(1-e^{cz_h^2})^2}\left(1-e^{-cz_h^2}\frac{I_2}{I_1}\right)I_1 \right]  \right|.
\end{eqnarray}

In this expression:
\[I_1=\int\limits_0^{z_h}(1+b\xi^2)^{3a}\xi^{1+\frac{2}{\nu}}d\xi, \]

\[I_2=\int\limits_0^{z_h}e^{c\xi^2}(1+b\xi^2)^{3a}\xi^{1+\frac{2}{\nu}}d\xi.\]

The result for the entropy density has the form:
 \begin{eqnarray}
s=\left(\frac{1}{z_h}\right)^{1+\frac{2}{\nu}}\frac{(1+bz_h^2)^{-3a}}{4G}, 
\end{eqnarray}
where $G$ is a dimensionless gravitational constant.

A characteristic relationship between baryon density and potential, independent of the anisotropy parameter $\nu$, is the following:
 \begin{eqnarray}\label{eq7}
\rho=-\frac{c\mu}{1-e^{cz_h^2}}.
\end{eqnarray}

For the numerical simulations, it is necessary to take into account the dependence of EoS on the baryonic potential. Therefore, the pressure in this model is calculated using the following formula:
 \begin{eqnarray}\label{eq8}
p=-\int\limits_0^{z_0}s\frac{dT}{dz_h}dz_h-\int\limits_0^{\mu_0}s\frac{dT}{d\mu}d\mu+\int\limits_0^{\mu_0}\rho d\mu.
\end{eqnarray}

The energy density can be found from the basic thermodynamic identity for relativistic hydrodynamics, combining the eqs. (\ref{eq4}--\ref{eq8}):
 \begin{eqnarray}
\varepsilon=-p+Ts+\rho \mu.
\end{eqnarray}

\section{Calibrating the model}

A parameter $c=0.227$~GeV$^2$, which appears in the initial metric form (1), is adjusted to the Regge spectra of the $\rho$ mesons based on the algorithm described in~\cite{10}. Parameters $a$ and $b$ are to be determined through a comparison of model predictions with lattice calculations. We decided to use the results of Ref.~\cite{11} because the calculations there are made with the physical masses of the $\rho$ mesons. This choice approximately corresponds to the physical case of quark masses. 

The calibration of holographic models in this work is achieved using of the method proposed in Ref.~\cite{12}. Within this method, dimensionless thermodynamic quantities are considered in the form of (\ref{eq4}--\ref{eq8}), with multiplication of them by a certain energy scale $L$ corresponding to conversion to units of GeV, and the gravitational horizon $z_h$ has units of GeV$^{-1}$. Notably, the power of GeV corresponding to a given physical quantity is determined by the power of the energy scale. Consequently, the scale $L$ will is determined alongside parameters $a$ and $b$.

The isotropic model is calibrated using the results of calculations of $\dfrac{s}{T^3}$ value, while the anisotropic case is adjusted to match the results of quark susceptibility $\dfrac{1}{T^2}\dfrac{\partial \rho_q}{\partial \mu_q}$. The quark potential is taken as $\mu_q=\dfrac{\mu_b}{3}$. This choice for the anisotropic model is related to the features of the phase behavior found in theoretical studies. The value of $\dfrac{s}{T^3}$ is unsuitable for the fitting procedure in this form  due to the occurrence of extreme temperature behavior in the anisotropic case, which exists at $\mu_b=0$. The relation $\rho_b(\mu_b)$ in formula (6) turns out to be independent of the parameter $\nu$, allowing it to be used for comparison with the lattice data. The parameters were adjusted using the least squares method, yielding the values shown in Table 1 (with the values of parameters $b$ and $c$ made dimensionless using the $L$ scale). The fit results for the isotropic and anisotropic models are shown in Figs. 1, 2.

The results of these fits exhibit lower quality compared to, for example, those in Ref.~\cite{13}.
However, since the model in this case is adjusted via the potential reconstruction method---where the dilaton potential is computed rather than predefined, and all free parameters reside in the initial metric---the model becomes more sensitive to minor parameter adjustments.
This results in an element of unpredictability with regard to the behavior of the chi-square function. In future work, we plan to implement more sophisticated parameter-tuning methodologies that leverage machine learning algorithms. Additionally, we will explore the impact of other deformation factors on the model's behavior to achieve better control over the calibration process.

\section{The results of numerical simulation of the QGP evolution
using the holographic equation of state}

 Considering the content of the papers, devoted to practical applications of lattice EOSs (for example,~\cite{14,15}), we conclude that further fine-tuning of the implemented equation of state is necessary--for instance, by matching it at low temperatures ($T$) with the hadron gas equation. 
  At the same time, we believe that a qualitative agreement between the simulation results and the calibrated lattice equations should be achieved, given that the holographic model aims to provide a complete description of the QCD phase diagram, including phase transitions.

As mentioned earlier, the complete application of EoS is possible only in combination with the equations of relativistic hydrodynamics.  Solving such a system is a highly complex task, so specialized software packages with dedicated numerical methods are typically employed. In this study, we use the relativistic hydrodynamics package MUSIC~\cite{16}, which is based on the MUSCL-type algorithm.  Among the available equations of state, we distinguish the ideal gas equation (for which explicit expressions exist) and tabulated lattice equations. We implement the tables for the holographic EoSs in full accordance with the specified lattice analogues, which allows the use of linear interpolation methods built into MUSIC to read the data.

It is also important that the MUSIC package is part of a larger software framework designed for step-by-step physical modeling of quark-gluon plasma evolution---from pre-equilibrium initial conditions to hadron transport models.  For this reason, our numerical simulations employ the iEBE-MUSIC package~\cite{17}, which uses MUSIC code as its hydrodynamic core. The following calculation scheme was used:
\begin{enumerate}
    \item The 3D MC Glauber package \cite{18}---a three-dimensional Glauber-type Monte Carlo model---generates the initial conditions for the subsequent hydrodynamical approach.
    \item The MUSIC package evolves the system using holographic equations of state until freeze-out at a specified energy density.  
    \item  The iSS package \cite{19}---performing event-by-event particle sampling from the freeze-out hypersurface--simulates particle emission while maintaining consistency between the hypersurface and transport model. 
    \item The UrQMD transport model \cite{20} handles the final stage of the simulation, including hadron re-scatterings and short-lived particle decays, providing with final hadron spectra output of the entire iEBE calculation in one of the standard UrQMD formats.
\end{enumerate}

A comparison of the corresponding predictions for the normalized by unity rapidity spectra of $\pi^{-}$-mesons in the kinematic region of the NA49 experiment for a wide range of impact parameters is shown in Fig. 3. Different lines correspond to the simulation results at the energy of $\sqrt{s} = 8.9$~GeV in iEBE-MUSIC for isotropic and anisotropic holographic equations, together with the results of the NEOS~\cite{21} equation of state, embedded in MUSIC. The latter case is taken as a reference. Data points from Ref.~\cite{22} correspond to the NA49 results for the least centrality class.  A similar comparison for central collisions with an impact parameter of $b < 2.5$ fm is shown in Fig. 4. Isotropic, anisotropic and NEOS EOSs are compared with the NA49 data for central collisions from Ref.~\cite{23}.

We observe that the rapidity distributions, up to a scaling factor, coincide across the different equations of state and also are in a good agreement with the experimental data. The overall scale in hydrodynamic simulations is highly sensitive to the details of the initial conditions, so it is not considered in this paper. These results demonstrate the successful calibration of all three equations of state while also highlighting the need for more advanced observables that exhibit greater sensitivity to the equation of state.

\section{Conclusion and outlook}

This paper presents a method for calibrating isotropic and anisotropic holographic equations of state in the ``light quarks'' limit.  The calibration leverages lattice QCD results with physical quark masses to enable numerical simulations of matter evolution under conditions relevant to real experiments. The holographic EOSs are integrated into the MUSIC relativistic hydrodynamics package. It is important to note that the computation time is comparable to that of built-in equations. Using the iEBE-MUSIC package for multi-stage simulations of QGP dynamics, we find that results obtained with holographic equations of state qualitatively match those from non-holographic equations.

Future work should focus on refining the hydrodynamic model parameters and optimizing the numerical implementation of the holographic EOS to better quantify its predictive accuracy.  Furthermore, the exploration of alternative holographic models and software products can facilitate the attainment of more precise conclusions regarding the nature of the simulation outcomes that depend on the equation of state employed.

\section*{FUNDING}
The authors acknowledge Saint-Petersburg State University for a research project 103821868.

\section*{CONFLICT OF INTEREST}
The authors declare that they have no conflicts of
interest.

\makeatletter
\renewcommand{\bibname}{REFERENCES} 
\renewcommand{\bibsection}{%
  \section*{\bibname}%
}
\makeatother


\bibliographystyle{maik}

\newpage

\begin{table}[h!]
\caption{Values of parameters obtained from the least squares method}
\begin{tabular}{|c|c|c|c|c|c|}
\hline
Model       & $\nu$ & $a$     & $b$     & $G$    & $L,$ GeV     \\ \hline
Isotropic   & 1     & 3.71  & 0.011 & 0.34 & 1.08  \\ \hline
Anisotropic & 4.5   & 3.949 & 0.034 & 0.81 & 1.01  \\ \hline
\end{tabular}
\end{table}

\newpage

\begin{figure}[h!]
\center
\includegraphics[scale=0.7]{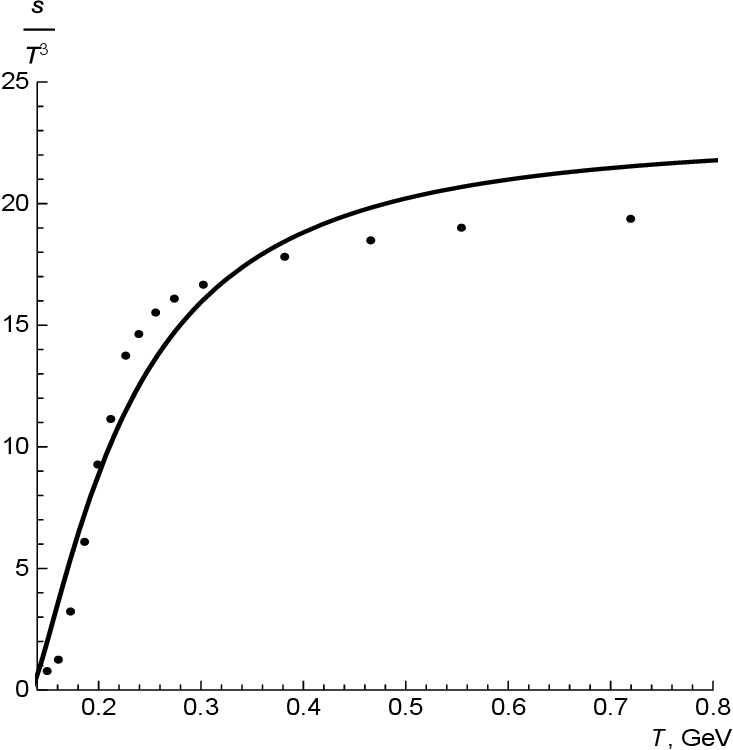}
\caption{Result of fitting ${s}/{T^3}$ ratio with the data from~\cite{19} for isotropic holographic model}
\end{figure}

\begin{figure}[h!]
\center
\includegraphics[scale=0.7]{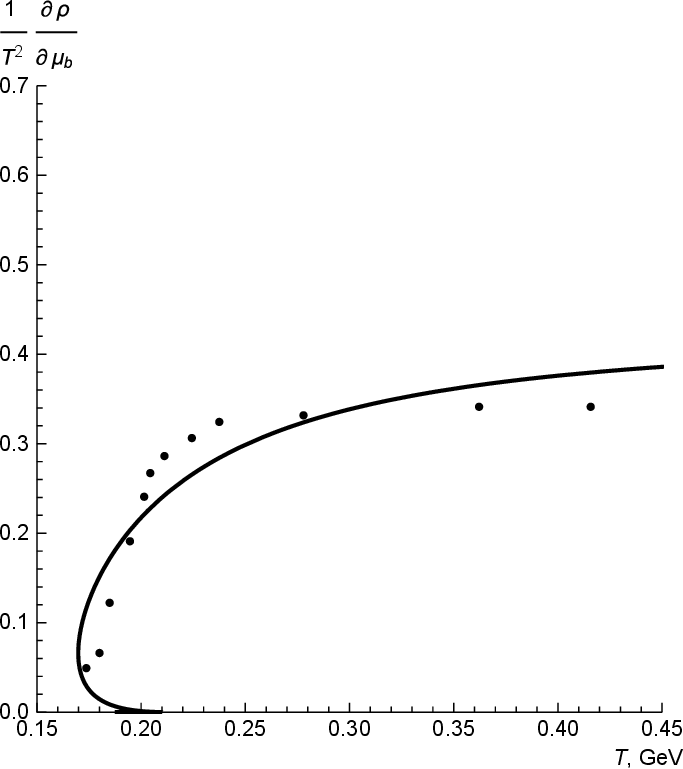}
\caption{Result of fitting normalized quark susceptibility with the data from ~\cite{19} for anisotropic holographic model}
\end{figure}

\newpage

\begin{figure}[h!]
\center
\includegraphics[scale=0.5]{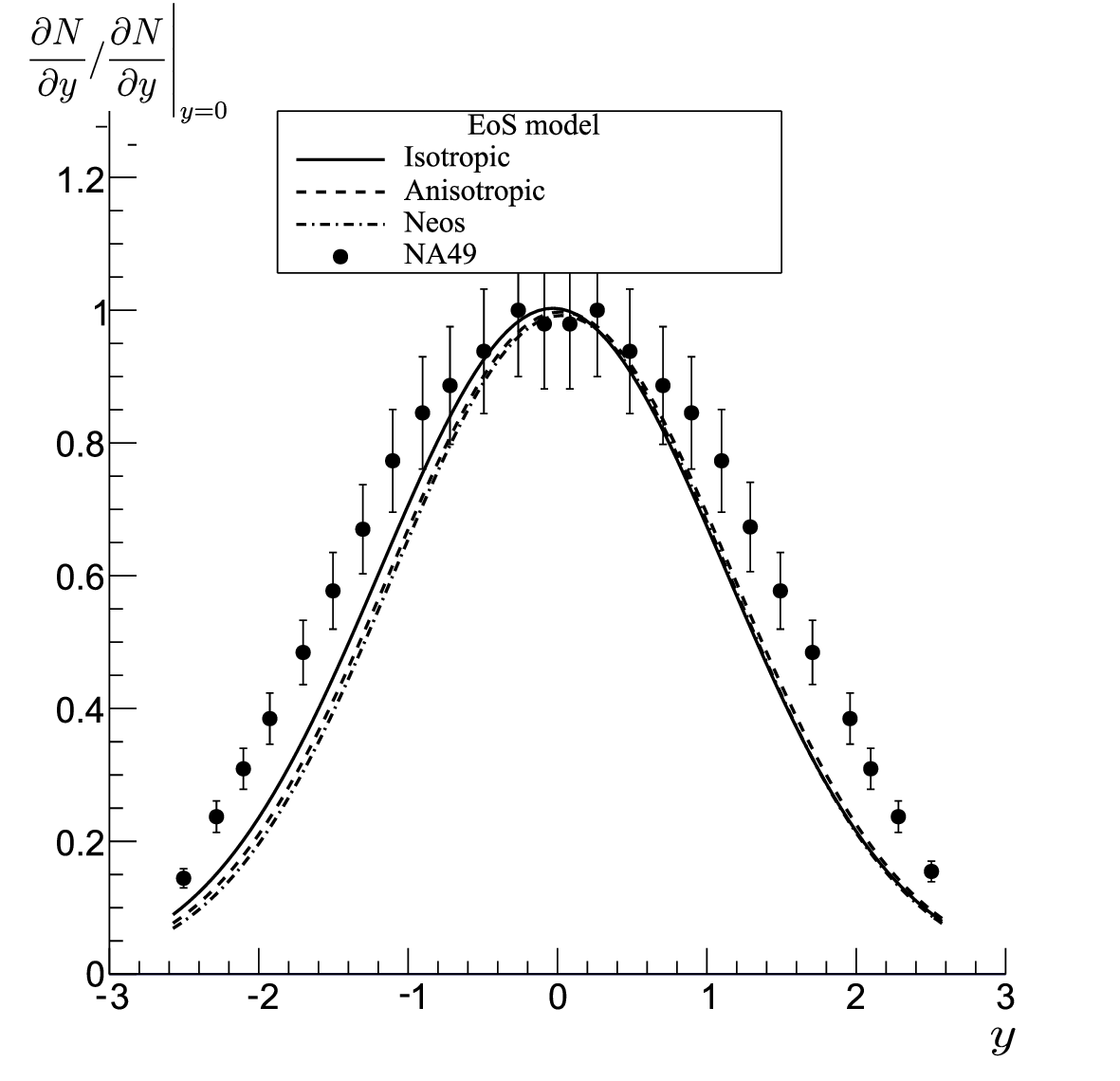}
\caption{ Normalized by unity y-spectra of $\pi^{-}$-mesons at $\sqrt{s}=8.9$ GeV and $b < 20$ fm for holographic EoSs and lattice NEOS EoS are compared with the data of NA49 for the highest centrality class from~\cite{22}}
\end{figure}

\begin{figure}[h!]
\center
\includegraphics[scale=0.5]{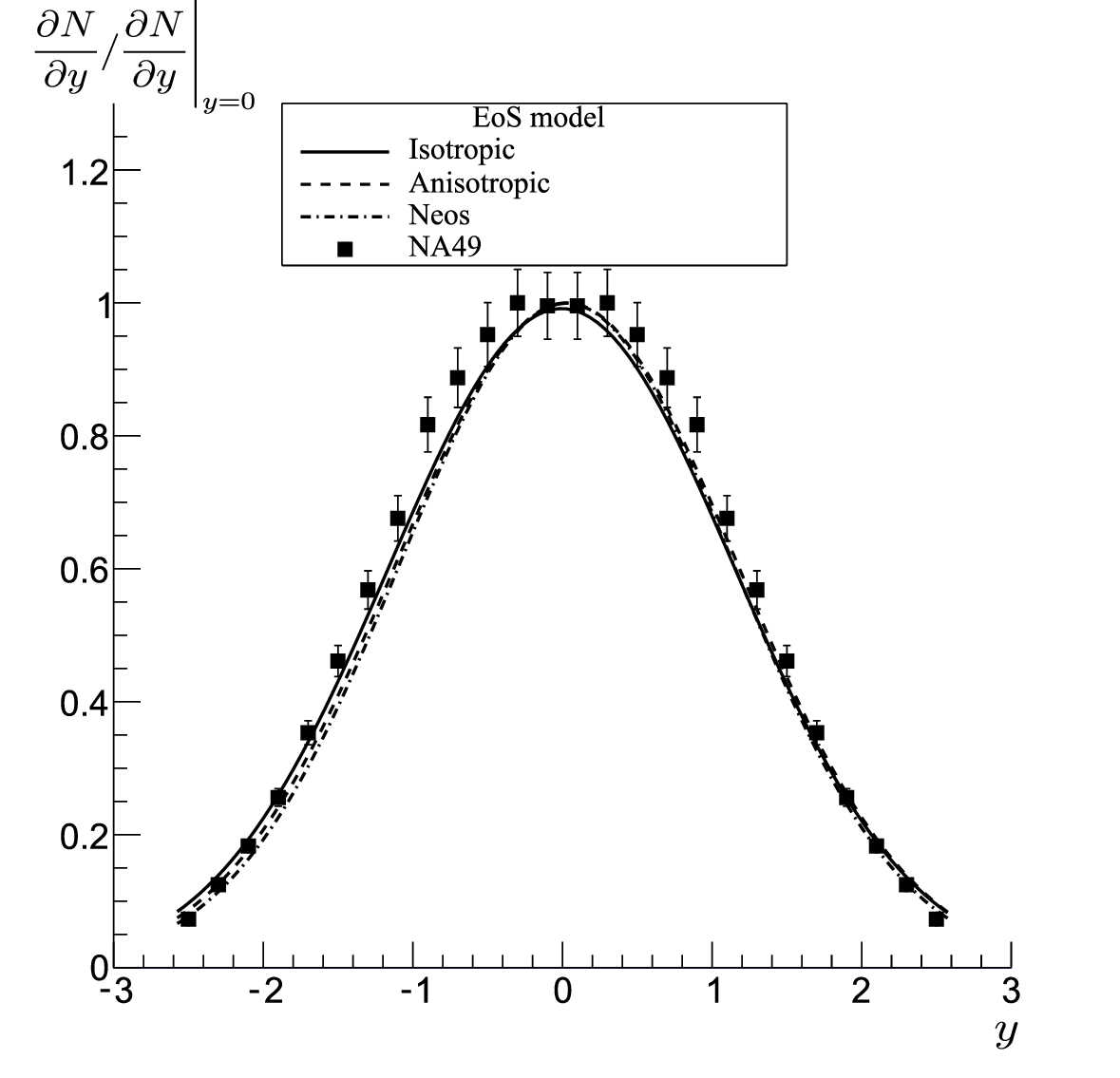}
\caption{Normalized by unity y-spectra of $\pi^{-}$-mesons at $\sqrt{s}=8.9$ GeV and $b < 2.5$ fm for holographic EoSs and lattice NEOS EoS are compared with the data of NA49 for the central collisions from~\cite{23}}
\end{figure}

\end{document}